\documentclass{PoS}
\usepackage{amsmath,amssymb,amsfonts,amscd,epsf,graphicx}

\title{Cutkosky Rules from Outer Space}

\ShortTitle{Cutkosky Rules from Outer Space}

\author{\speaker{Dirk Kreimer}%
        \thanks{The results presented here grew out of a collaboration with Spencer Bloch. Also, many thanks to David Broadhurst for stimulating discussions on the subject.}\\
       Humboldt Univ.\\
       E-mail: \email{kreimer@physik.hu-berlin.de}}

\abstract{We overview recent results on the mathematical foundations of Cutkosky rules. We emphasize that the two operations of shrinking an internal edge or putting internal lines on the mass-shell are natural operation on the cubical chain complex studied in the context of geometric group theory. This together with Cutkosky's theorem regarded as a theorem which informs us about variations connected to the monodromy of Feynman amplitudes allows for a systematic approach to normal and anomalous thresholds, dispersion relations and the optical theorem. In this report we
follow \cite{BK} closely.}

\FullConference{Loops and Legs in Quantum Field Theory\\
         24-29 April 2016\\
         Leipzig, Germany}

         \newcommand{\be}{\begin{equation}}
\newcommand{\ee}{\end{equation}}
\newcommand{\bea}{\begin{eqnarray}}
\newcommand{\eea}{\end{eqnarray}}
\newcommand{\beas}{\begin{eqnarray*}}
\newcommand{\eeas}{\end{eqnarray*}}

\def\cubical{{\;\raisebox{10mm}{\epsfysize=80mm\epsfbox{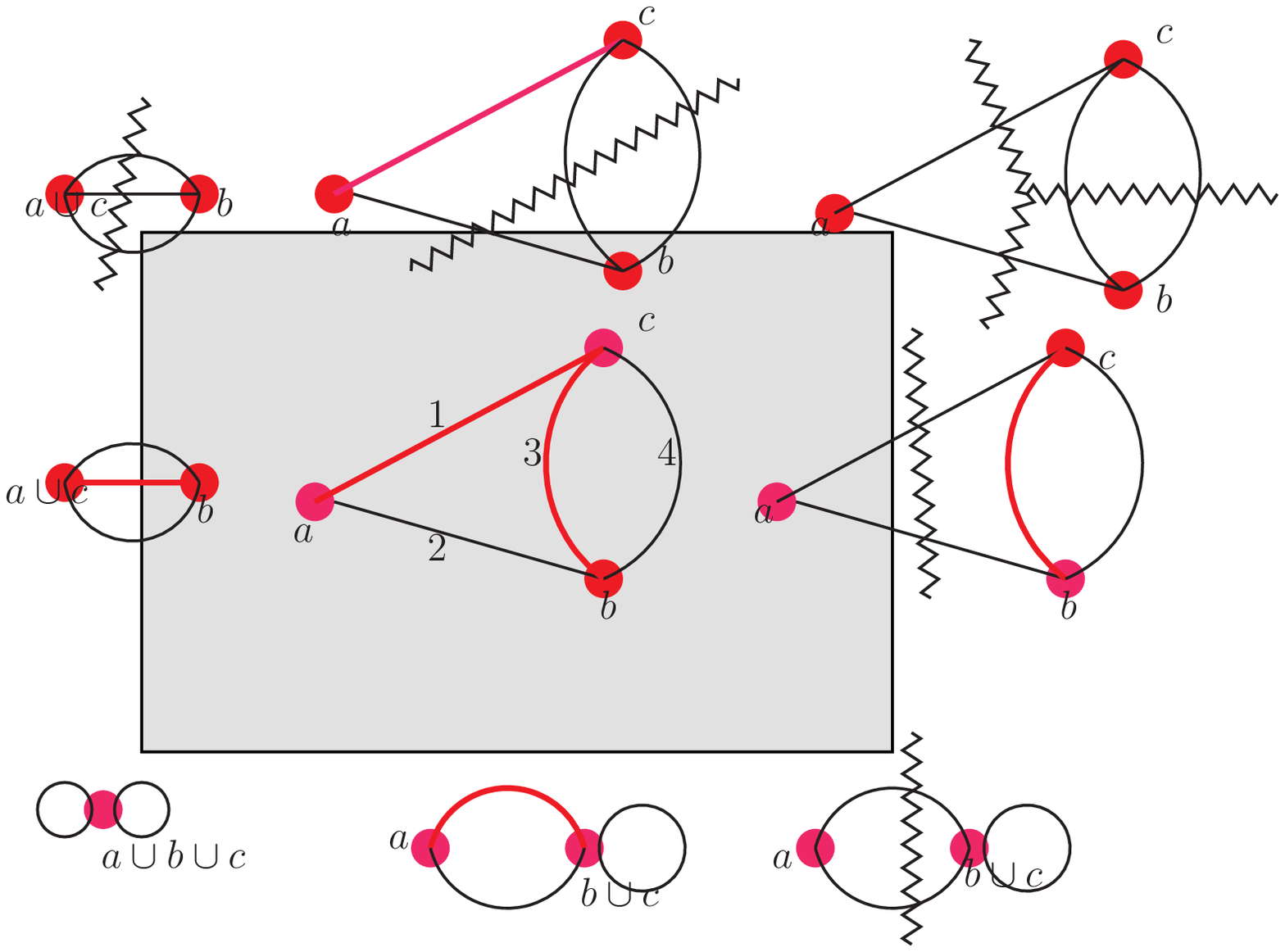}}\;}}
\def\cubicalM{{\;\raisebox{10mm}{\epsfxsize=140mm\epsfbox{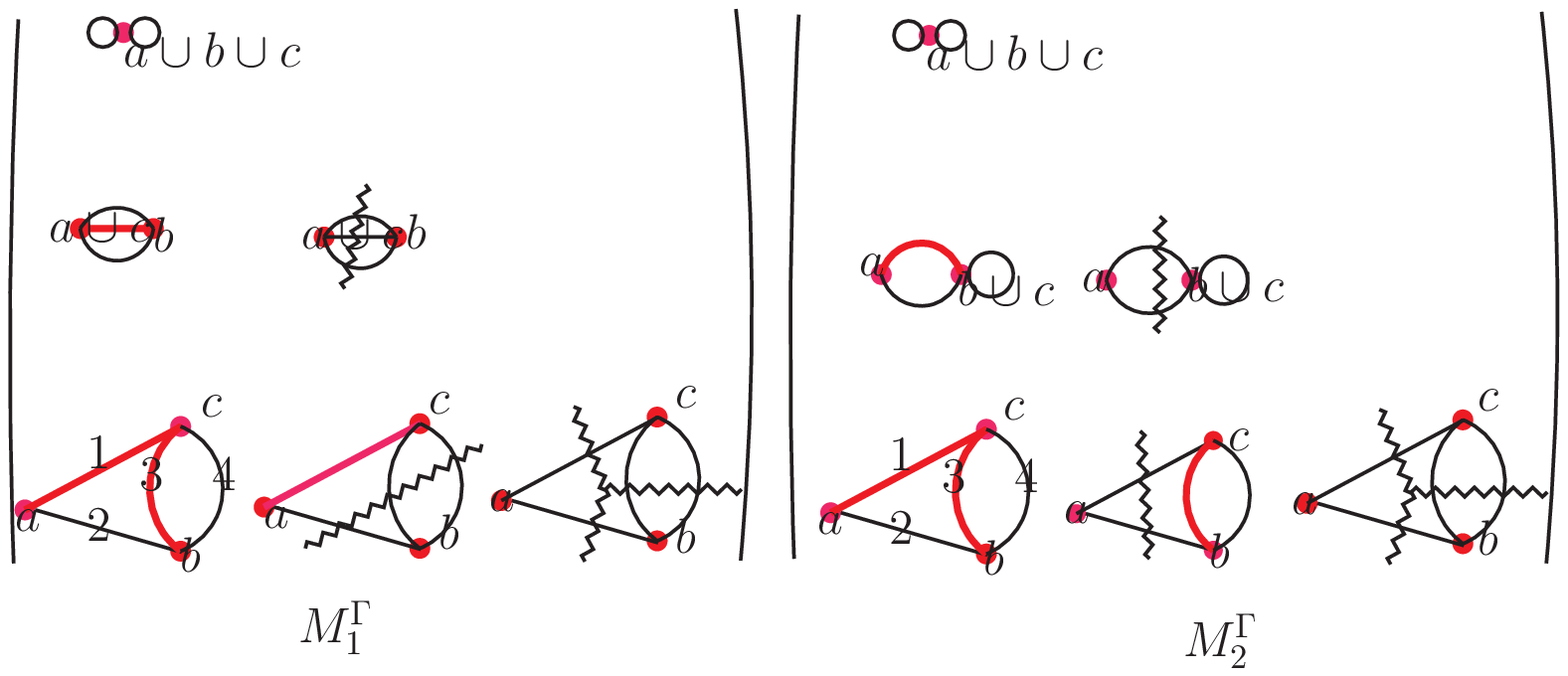}}\;}}

\def\trianglematrix{{\;\raisebox{-40mm}{\epsfxsize=80mm\epsfbox{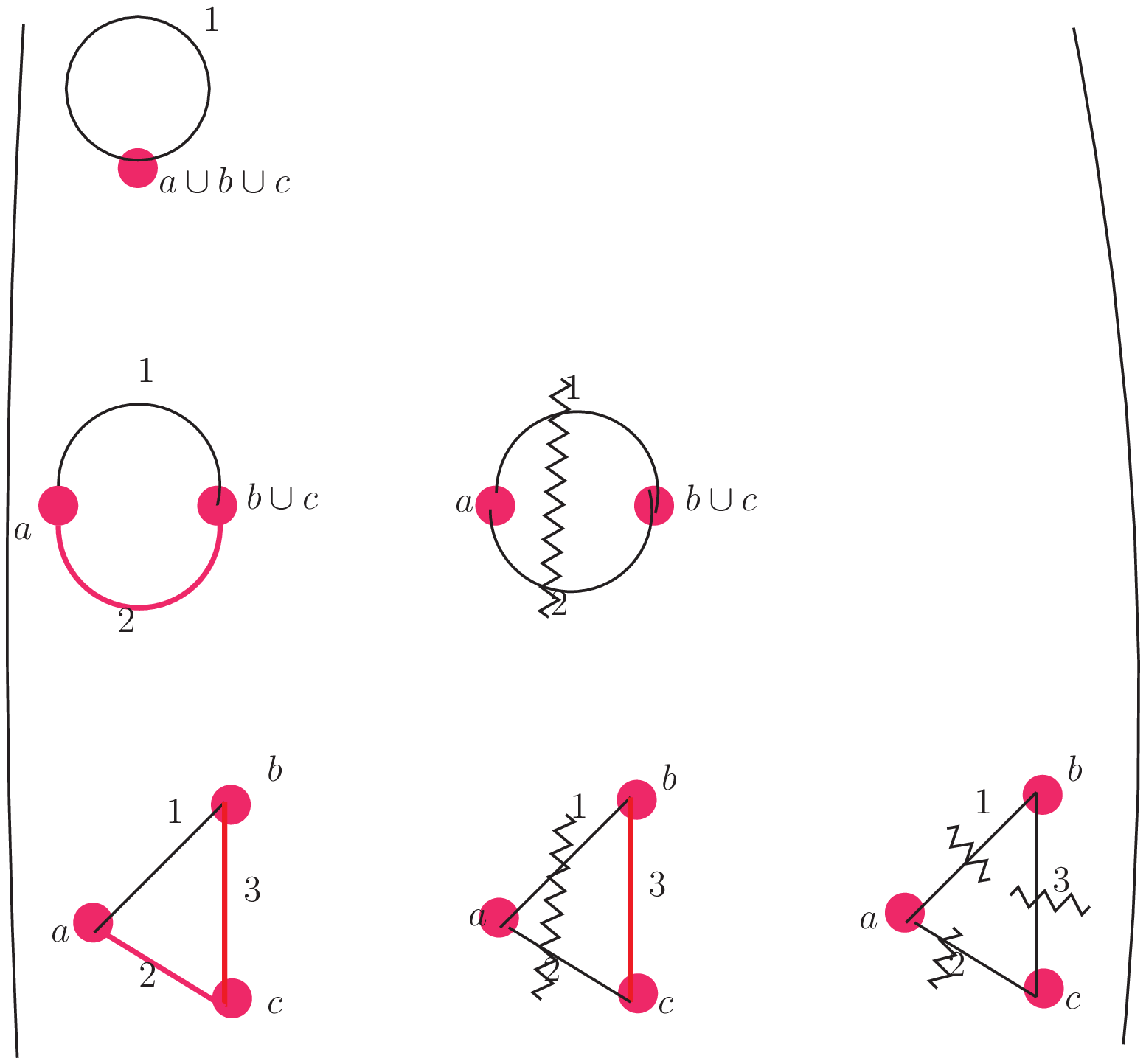}}\;}}
\def\trianglecubical{{\;\raisebox{-40mm}{\epsfxsize=80mm\epsfbox{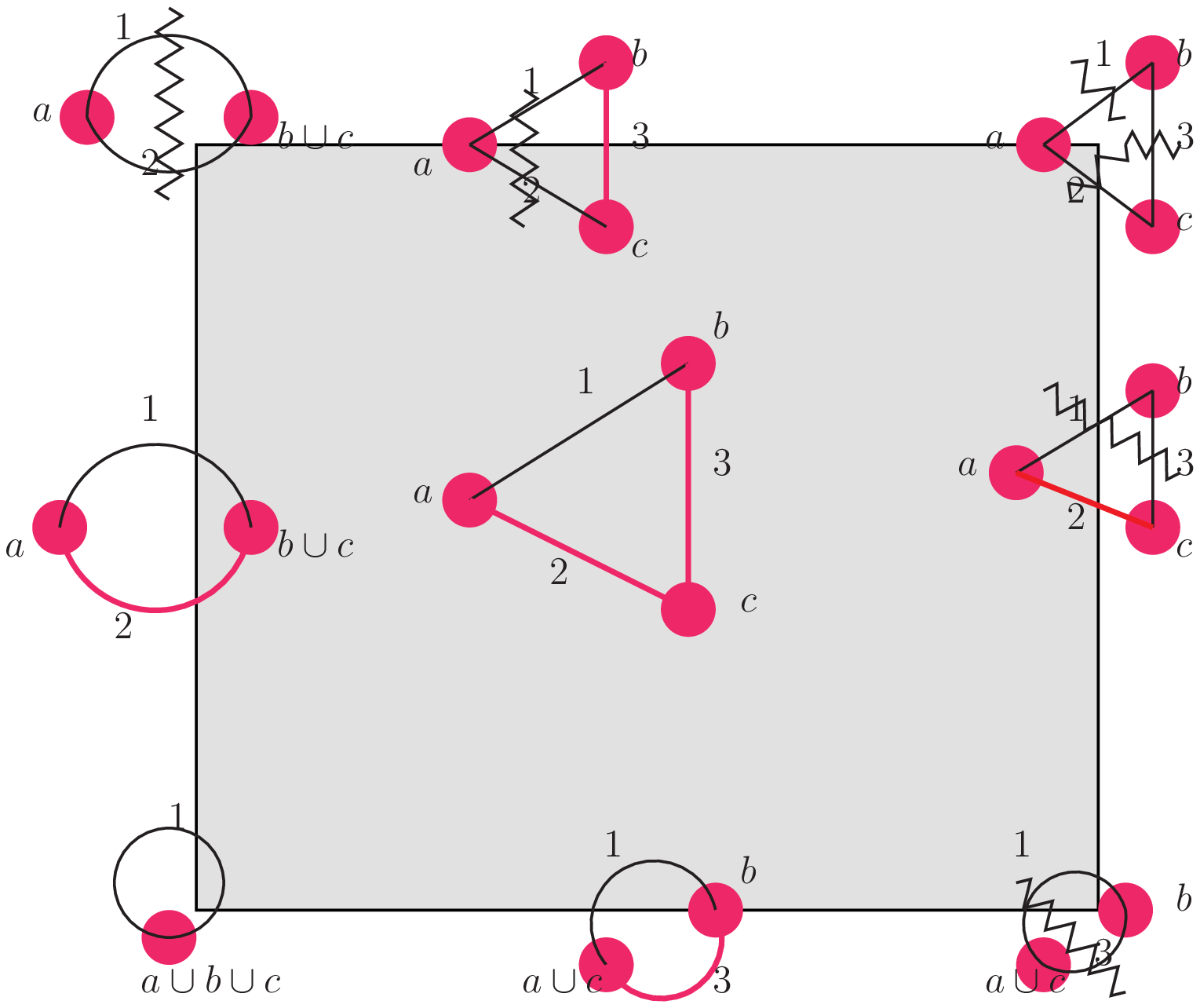}}\;}}

\begin{document}

\section{Motivation}
Understanding of the analytic structure of the contribution of a graph to a Feynman amplitude, a time-honored problem \cite{Pham},
is related to an analysis of
its reduced graphs and the graphs in which internal edges are on the mass-shell. The former case relates to graphs in which  internal edges shrink. The latter case relates to graphs with cut edges. The set of cut edges is uniquely determined by the choice of a spanning forest for the graph:
such a spanning forest defines a unique set of edges connecting distinct components of the forest. It is those edges we will put on the mass-shell.

Pairs of graphs and a chosen ordered  spanning tree or forest deliver the cubical chain complex \cite{HV}.

A given ordering of the edges of the spanning tree $T$ defines a sequence of spanning forests $F$, and to any pair $(\Gamma,F)$ for fixed $\Gamma$ 
we can associate:\\ 
-a reduced graph $\Gamma_F$ obtained by shrinking all edges of $\Gamma$ to length zero which do not connect different components of the spanning forest\\ 
-a cut graph $\Gamma^F$ where all those edges connecting different components are put on-shell, so are marked by a Cutkosky cut,\\
-the set of graphs $G^F=\Gamma-E_{\Gamma_F}$ obtained from $\Gamma$ by removing the edges which connect distinct components of the spanning forest. 

Such data define a cell-complex. With it they define a set of lower triangular matrices, one for each ordering of the edges in $T$,  which allow to analyse a graph amplitude from its reduced graphs and the variations obtained by putting internal edges on-shell.   
\subsection{Results}
A sequence of cuts (edge sets $\epsilon_i$ determines from $i$-component forests, $i\geq 2$)
\[\epsilon_2\to \epsilon_3\to\cdots\to \epsilon_{v_\Gamma}\] 
will shift the normal threshold $s_0(\epsilon_2)$ associated with a chosen cut $\epsilon_2$ to anomalous thresholds
\[
s_0(\epsilon_2)\to s_1(\epsilon_3)\to\cdots\to s_{v_\Gamma-2}(\epsilon_{v_\Gamma}).
\]

The resulting sequence of anomalous thresholds $s_i(\epsilon_{i+2})$, $i>0$ is a sequence of values for a channel variable $s$ defined by $\epsilon_2$.
They are computed from the divisors associated to $\epsilon_{i+2}$. The latter are functions of all kinematical variables.
For example, for the one-loop triangle the divisor in $\mathbb{C}^3$ associated to $\epsilon_3$ is a simple function of 
\[
\lambda(p_1^2,p_2^2,p_3^2)=p_1.p_2^2-p_1^2p_2^2=p_2.p_3^2-p_2^2p_3^2=p_3.p_1^2-p_3^2p_1^2,\,p_1+p_2+p_3=0.
\] 
The three representations of $\lambda(p_1^2,p_2^2,p_3^2)$ allow to compute $s_1(\epsilon_3)$ for the three choices of a channel variable $s=p_3^2$ or $s=p_1^2$ or $s=p_2^2$ respectively.

As a result, to a graph $\Gamma$  we can assign a collection of lower triangular matrices $M_i^\Gamma$ with the following properties:\\
i) All entries in the matrix correspond to well-defined integrable forms under on-shell renormalization conditions.\\
ii) Anomalous thresholds $s_i$ are determined from properties of graph polynomials. They provide lower boundaries for dispersion integrals associated to these integrable forms.\\ 
iii) Along the diagonal in the matrices $M_i^\Gamma$ we find leading threshold entries: all quadrics for all edges in a graph  are on the mass-shell.\\
iv)  The variation of a column in $M_i^\Gamma$ wrt to a given channel is given by the column to the right.\\ 
v) The subdiagonal entries $(M_i^\Gamma)_{k,k-1}$ are determined from the diagonal entries  $(M_i^\Gamma)_{k-1,k-1}$
and $(M_i^\Gamma)_{k,k}$ via a dispersion integral. This gives $(k-1)$ two-by-two matrices each of which has an interpretation via the optical theorem.
This hence determines the first subdiagonal.\\
vi)  Continuing, all subdiagonals and hence the whole matrix $(M_i^\Gamma)_{r,s}$ is determined via iterated dispersion.
This answers the question how to continue the optical theorem beyond two-point functions.
\section{The cubical chain complex}
We follow  \cite{HV}.
Consider a pair $(\Gamma,T)$ of a bridge free graph $\Gamma$ and a chosen spanning tree $T$ for it.
Assume $T$  has $k$ edges. Consider the $k$-dimensional unit cube. It has origin $(0,\cdots,0)$ and $k$ unit vectors
$(1,0,\cdots,0),\ldots$, $(0,\cdots,0,1)$ form its edges regarded as $1$-cells. A change of ordering of the edges of $T$ permutes those edges.

The origin is decorated by a rose on $|\Gamma|$ petals, and the corner $(1,1,\cdots,1)$ decorated by $(\Gamma,V_\Gamma)$, with $k=|V_\Gamma|-1$,
and we regard $V_\Gamma$ as a spanning forest consisting of $k+1$ distinct vertices.

The complex is best explained by assigning graphs as in the following example.
$$\cubical$$
The cell is two-dimensional as each of the five spanning trees of the graph $\Gamma$, the dunce's cap graph, in the middle of the cell has length two.

We have chosen a spanning tree $T$ provided by the edges $e_1$ and $e_3$, indicated in red. The boundary of our two-dimensional cell has four one-dimensional edges, bounded by two of the four 0-dimensional corners each.

To these lower dimensional cells we assign graphs as well as indicated.

The spanning tree has length two and so there are $2=2!$ orderings of its edges, and hence two lower triangular $3\times 3$ matrices  $M^\Gamma_i$ which we can assign to this cell.

They look as follows:
$$\cubicalM $$
These square matrices are lower triangular. 
Note that we have cuts which separate the graph into two components determining a normal threshold which appears already in a reduced graph on the diagonal, and in the lower right corner a cut into three components, which determines an anomalous threshold. All these cuts determine variations, as stated in Cutkosky's theorem \cite{BK}.

\section{Cutkosky's theorem}
We quote from \cite{BK} where you find details. For a graph $\Gamma$ and a choosen spanning forest $F$ we let the quotient graph $\Gamma''$ -the reduced graph-
be the graph obtained by shrinking all edges $e\in E'$ of $\Gamma$ which do not connect  distinct components of $F$, so $E'=E_\Gamma-E''$,
and $E''$  all edges of $\Gamma$ which do connect distinct components of $F$.  

Assume the reducedgraph $\Gamma''$ has a physical singularity  at an external momentum point $p''$, i.e. the intersection $\bigcap_{e\in E''}Q_e$ of the propagator quadrics associated to edges in $E''$ has such a singularity at a point lying over $p''$. Let $p$ be an external momentum point for $\Gamma$ lying over $p''$. Then the variation of the amplitude $I(\Gamma)$ around $p$ is given by Cutkosky's formula
\[\text{var}(I(\Gamma)) = (-2\pi i)^{\# E''}\int\frac{\prod_{e\in E''} \delta^+(\ell_e)}{\prod_{e\in E'} \ell_e}.
\]

\section{Anomalous thresholds}
Let us come back to a generic graph $\Gamma$. We want to determine anomalous thresholds. With their help, dispersion relations can be established when real analycity in kinematical variables can be established.

We analyse the Landau singularities of $\Gamma$ in terms of $\Gamma/e$, where
$e$ is such an edge. To completely analyse the graph, we have to consider all possibilities to shrink it edge after edge (the generalization to multiple edges is in \cite{BK}). 

We have for the second Symanzik polynomial $\Phi$
\[ 
\Phi(\Gamma)=\overbrace{\Phi(\Gamma/e)}^{=:X}+A_e\overbrace{\left\{ \Phi(\Gamma-e)-m_e^2\psi(\Gamma/e) \right\} }^{=:Y}-A_e^2\overbrace{m_e^2\psi(\Gamma-e)}^{Z}.
\]
Solving $\Phi(\Gamma)=0$ for $A_e$ is a quadratic equation with coefficients $X,Y,Z$. Note that $Z>0$ is independent of kinematical variables, while $X,Y$ depend on momenta and masses. In particular, for a chosen channel variable $s$ we can write $X=sX_s+N$, with $X_s$ independent of kinematics and $N$ a constant in the channel variabel $s$. It depends on other kinematic variables though.
In terms of parametric variables, $X,Y$ are functions of the parametric variables of the reduced graph.

The above quadratic equation has a discriminant $D=Y^2+4XZ$, and we find a physical Landau singularity for positive $Y$ and vanishing discriminant $D=0$. Define $Y_0:=Y(p_A^{\Gamma/e},\{Q,M\})$ to be the evaluation of $Y$ by evaluating  parametric variables at the point of the Landau singularity for the reduced graph.

The condition $D=0$ allows to determine the anomalous threshold from
\[
s(\{A\},\{Q,M\})=\frac{Y^2-4ZN}{4ZX_s},
\]
minimizing over parametric variables $A_e\geq 0$.

Let $\mathit{T}_s^\Gamma$ be the set of all ordered spanning trees $T$ of a fixed graph $\Gamma$ which allow for the same associated channel variable $s$. 

We have the following result.\\
i) A necessary and sufficient condition for a physical Landau  singularity is $Y_0>0$ with $D=0$.\\
ii)  The corresponding anomalous threshold $s_F$ for fixed masses and momenta $\{M,Q\}$ is given as the minimum of $s(\{a,b\},\{Q,M\})$ varied over edge variables $\{a,b\}$. It is finite ($s_F>-\infty$) if the minimum is a point inside $p\in\mathbb{P}^{e_\Gamma-1}$ in the interior of the 
integration domain $A_i>0$. If it is on the boundary
of that simplex, $s_F=-\infty$.\\
iii) If for all $T\in \mathit{T}_s^\Gamma$ and for all their forests $(\Gamma,F)$ we have $s_F>-\infty$, the Feynman integral $\Phi_R(\Gamma)(s)$ is real analytic as a function of $s$
for $s<\min_F\{s_F\} $.\\
iv) For $Y>0$ and $X<0$, both zeroes of $\Phi(\Gamma)=0$  appear for $A_e>0$. For $Y>0$ and $X>0$, only one zero is inside the domain of integration. As a result for $X=0$ corresponding to the threshold provided by the reduced $\Gamma/e$  we have a discontinuity. 

\section{Example}
We consider the triangle graph $\Delta$. 
In fact, we augment it with one of its three possible spanning trees, say on edges $e_2,e_3$, so $E_T=\{e_2,e_3\}$. The corresponding cell in the cubical chain complex is
\be \label{trianglecell}
\trianglecubical
\ee

For the  Cutkosky cut we choose two of the three edges, say $\epsilon_2=\{e_1,e_2\}$. This defines the channel $s=p_a^2$ and the matrix $M^\Delta_1$.

The  other cut in that matrix is the full cut  separating all three vertices.

We give $M^\Delta_1$ in the following figure:
$$
M^\Delta_1=\trianglematrix
$$
We now calculate:
\beas
\Phi_\Delta & = & \overbrace{p_a^2A_1A_2-(m_1^2A_1+m_2^2A_1)(A_1+A_2)}^{=\Phi_{\Gamma/e_3}}\\
 & & +A_3((p_b^2-m_3^2-m_1^2)A_1+(p_c^2-m_1^2-m_3^2)A_2)-A_3^2m_3^2,
\eeas
so 
\beas 
\Phi_\Delta & = & \Phi_{\Delta/e_3}
  +A_3Y-A_3^2 m_3^2\overbrace{\psi_{\Delta-e_1}}^{=1},
\eeas
as announced:
\[
X=\Phi_{\Delta/e_3},\,Y=\overbrace{(p_b^2-m_3^2-m_1^2)}^{=:l_1}A_1+\overbrace{(p_c^2-m_1^2-m_3^2)}^{=:l_2}A_2,\,Z=m_3^2.
\]
We have $Y_0=m_2 l_1+m_1 l_2$, and need $Y_0>0$ for a Landau singularity.

Solving $\Phi(\Delta/e_3)=0$ for a Landau singularity determines
the familiar physical threshold in the $s=p_a^2$ channel, leading for the reduced graph to
\be
p_Q: s_0=(m_2+m_3)^2,\, p_A: A_1m_1=A_2m_2. 
\ee

We let $D=Y^2+4XZ$ be the discriminant. For a Landau singularity we need 
\[
D=0.
\]
We have
\be
\Phi_\Delta=-m_3^2\left(A_3-\frac{Y+\sqrt{D}}{2m_3^2}\right)\left(A_3-\frac{Y-\sqrt{D}}{2m_3^2}\right),
\ee
where $Y,D$ are functions of $A_1,A_2$ and $m_1^2,m_2^2,m_3^2,s,p_b^2,p_c^2$. Note that at $D=0$ we have
\[
2m_3^2A_3=A_1l_1+A_2l_2,
\]
which determines a co-dimension one (a line) hypersurface of $\mathbb{P}^2$. Finding the anomalous thresold determines a point on this line (it fixes the ratio $A_1/A_2$), and hence the anomalous threshold determines a point in $\mathbb{P}^2$.
We can write
\[
0=D=Y^2+4Z(sA_1A_2-N),
\] 
with $N=(A_1m_1^2+A_2m_2^2)(A_1+A_2)$ $s$-independent.

This gives
\be 
s(A_1,A_2)=\frac{4ZN-(A_1l_1+A_2l_2)^2}{4ZA_1A_2}=:\frac{A_1}{A_2}\rho_1+\rho_0+\frac{A_2}{A_1}\rho_2.
\ee

Define two Kallen functions $\rho_1=-\lambda_1=-\lambda(p_b^2,m_1^2,m_3^2)$ and
$\rho_2=-\lambda_2=-\lambda(p_c^2,m_2^2,m_3^2)$. Both are real and non-zero off their threshold or pseudo-threshold.

Then, for 
\[
\rho_1>0,\,\rho_2>0,
\]
we find the threshold $s_1$ at
\be 
s_1=(m_1+m_2)^2+\frac{4m_3^2(\sqrt{\lambda_2}m_1-\sqrt{\lambda_1}m_2)^2-(\sqrt{\lambda_1}l_2+\sqrt{\lambda_2}l_1)^2}{4m_3^2\sqrt{\lambda_1}\sqrt{\lambda_2}}.
\ee

On the other hand for the coefficients of $\rho_1<0$ and/or $\rho_2<0$  we find a minimum
\be s_1=-\infty,
\ee
along the boundaries  $A_1=0$ or $A_2=0$.

The domains $Y>0,X<0$ and $Y>0,X>0$ determine the domains of parametric integration for the variation prescribed by Cutkosky's theorem,
whilst the normal and anomalous threshold (when finite)  determine the lower boundaries of the dispersion integrals needed to reconstruct the function from its variation.

Let us now discuss the triangle in more detail.
It allows three spanning trees on two edges each, so we get six matrices $M^\Delta_i$, $i=1,\ldots,6$ altogether,
by having two possibilities to order the two edges for each spanning tree.

The six matrices $M^\Delta_i$ come in groups of two for each spanning tree.

For each of the three spanning trees we get a cell as in (\ref{trianglecell}).

The boundary operator for such a cell in the cubical cell complex of \cite{HV} is the obvious one stemming from co-dimension one hypersurfaces at $0$ or $1$
with suitable signs. So the square populated by the triangle $\Delta$ in (\ref{trianglecell}) has four boundary components, the edges populated by the four graphs as indicated. Those four edges are the obvious boundary of the square. 

If we now consider all graphs in (\ref{trianglecell}) as evaluated by the Feynman rules, we can consider for a given cell a boundary operator which replaces evaluation at 
the $x_e=0$-hypersurface by shrinking edge $e$, and evaluation at the $x_e=1$-hypersurface by setting edge $e$ on the mass-shell.

Then, to check that this is a boundary operator for the amplitudes defined by the graphs in (\ref{trianglecell}) we need to check that the amplitudes for the  four graphs at the four corners are uniquely defined from the amplitudes of the graphs at the adjacent edges: for example, the imaginary part of the amplitude of the  graph on the left vertical edge is related to the amplitude of the graph at the upper left corner:
This imaginary part must be also obtained from  shinking  edge $e_3$ in the graph on the upper horizontal edge by setting $A_3$ to zero in the integrand and integrating over the hypersurface $A_3=0$ of the integration simplex $\sigma_\Delta$. This is indeed the case, and similar checks work for all other corners.

In summary, the analytic structure of Feynman amplitudes realized the structure of the cubical chain complex. The latter is highly non-trivial.
Its further study in the conext needed for physics  will inform our understanding of amplitudes considerably. Future work will be dedicated in understanding the relation between the monodromy of physical singularities and the fundamental group underlying Outer Space as used in \cite{HV}.


\begin{thebibliography}{99}
 \bibitem{BK}
 S.\ Bloch, D.\ Kreimer,
  {\em Cutkosky rules and Outer Space},
  arXiv:1512.01705 [hep-th].
  \bibitem{Pham}
  F.\ Pham, 
  {\em Introduction \`a l'Etude Topologique des Singularit\'es de Landau},
Gautheir-Villars (1967).
\bibitem{HV}
A.\ Hatcher, K.\ Vogtmann,
{\em Rational Homology of Aut($F_n$)}, Math. Research
Lett. 5 (1998) 759-780.
\end{thebibliography}
\end{document}